\documentclass[aps,prd,onecolumn,showpacs,floats,floatfix,letterpaper,nofootinbib,superscriptaddress,notitlepage]{revtex4-1}
\pdfoutput=1
\usepackage{hyperref}
\usepackage{amssymb,amsmath,latexsym,mathrsfs}
\usepackage{graphicx,subfigure}
\usepackage{epsfig}
\usepackage{varioref,xr-hyper}
\usepackage{color}
\usepackage{multirow}
\usepackage{array}
\hypersetup{
     colorlinks   = true,
     citecolor    = blue,
     linkcolor = magenta
     }
\begin{document}
\title{Redshift drift in an inhomogeneous universe: averaging and the backreaction conjecture}
\author{S. M. Koksbang}
\email{koksbang@phys.au.dk}
\affiliation{Department of Physics and Astronomy, Aarhus University, 8000 Aarhus C, Denmark}
\author{S. Hannestad}
\affiliation{Department of Physics and Astronomy, Aarhus University, 8000 Aarhus C, Denmark}
\begin{abstract}
An expression for the average redshift drift in a statistically homogeneous and isotropic dust universe is given. The expression takes the same form as the expression for the redshift drift in FLRW models. It is used for a proof-of-principle study of the effects of backreaction on redshift drift measurements by combining the expression with two-region models. The study shows that backreaction can lead to positive redshift drift at low redshifts, exemplifying that a positive redshift drift at low redshifts does not require dark energy. Moreover, the study illustrates that models without a dark energy component can have an average redshift drift observationally indistinguishable from that of the standard model according to the currently expected precision of ELT measurements.
\newline\indent
In an appendix, spherically symmetric solutions to Einstein's equations with inhomogeneous dark energy and matter are used to study deviations from the average redshift drift and effects of local voids. 
\end{abstract}

\maketitle

\section{introduction}
Modern cosmology revolves around the dark energy problem - the fact that the standard model of cosmology presumes that the present day energy density of the Universe is dominated by a substance known as dark energy of which the physical origin is entirely unknown. A large number of theories have been proposed as the physical foundation for dark energy and several of the resulting cosmological models are not falsifiable by current observations. Constraining and eventually excluding physical models of dark energy with future observations is a top priority goal in cosmology.
\newline\indent
Different physical models of dark energy generally lead to different overall dynamics of the Universe. Hence, observables that yield information about the expansion history of the Universe, described by the Hubble parameter $H(z)$, are important for reaching this goal. One such observable is the redshift drift $\delta z$. This quantity describes the minute temporal changes in the redshift of a comoving luminous source that take place because of acceleration or deceleration of the Universe's expansion. The fact that such a drift in the redshift should in principle occur in any non-coasting universe was first realized in the 1960's \cite{Sandage, McVittie}. It was initially deemed futile to measure the redshift drift since it would require time intervals of the order $10^7$ years with the technology available at that time. It was later \cite{Loeb} proposed that more modern technology might allow redshift drift measurements by using {\em e.g.} the Lyman $\alpha$ forest at high redshifts ($z \gtrsim 2.5$) as observable. Elaborating on this suggestion, it is today estimated that the next generation of extremely large telescopes (ELTs) can be used to measure the redshift drift of absorption lines of QSO spectra in the redshift range $2\lesssim z\lesssim 5$ using time intervals of the order $10 - 100$ years \cite{sigma, realtime, drift_test_ldcm3, codex} \footnote{Redshift drift measurements at lower redshifts using Square Kilometer Array (SKA) observations of galaxies may also be feasible in the near future \cite{drift_with_SKA, drift_observation_made}. This will not be considered further here.}. Thus, measuring the redshift drift has become a practicable possibility
\footnote{In fact, redshift drift has in principle already been measured \cite{drift_observation_made}!}.
\newline\newline
The redshift drift in FLRW models is given by the expression:
\begin{equation}\label{eq:zdrift_flrw}
\delta z = \left[ (1+z)H_0 - H(t_e)\right] \delta t_0
\end{equation}
Subscripts $0$ denote time of observation (present time) while the subscript $e$ denotes evaluation at the time of emission. $\delta t_0$ is the time interval between two measurements of the redshift of a given source.
\newline\indent
The formula shows that if the Hubble constant is known, {\em e.g.} through low-redshift supernova observations, then measurements of the redshift drift directly yield $H(z)$ for any cosmological model with an FLRW metric. Thus, $\delta z$ is highly dependent on the expansion history of the Universe and it is the hope that future measurements of $\delta z$ over large redshift intervals will put significant constraints on the expansion history of the Universe and hence on possible solutions to the dark energy problem. It has, for instance, over the last couple of decades become clear that a large local void with us near its center can lead to redshift-distance relations equal to those predicted by the standard model but without introducing dark energy (see {\em e.g.} \cite{Havard1, Havard2, Troels_supernova, bolejko_void, wessel_void,   Dallas_mexico_void, minimum_void, February, enqvist, dallas_short, LTB_new_Sundell}). It turns out that such models generally have redshift drifts that deviate significantly from that of the standard model \cite{distinguish_with_drift, Copernican_drift, analytic_LTB, drift_ltb_vs_DE, drift_LTB, void_vs_DE}. One difference which has gotten particularly much attention is that the redshift drift is negative in LTB void models for all redshifts \cite{drift_LTB} while it is positive at small redshifts in the standard model. Redshift drift measurements should thus make it possible to exclude either the standard model or the cosmological models based on LTB models with us at the center of a large void mimicking accelerated expansion. It should be noted that a large local void does not exclude a positive redshift drift at low redshifts; if the given model includes a dark energy component, the redshift drift may be positive at low redshifts regardless of the existence of a local void. The possible existence of such a local void is important {\em e.g.} for dark energy phenomenology and parameter inferences based on standard observables \cite{coldspot, Mikko3, Mikko2, residual_hubblebubble, fake_de, localStructure_omegaL, localStructure_omegaDE}.
\newline\indent
Aside from in local void models, redshift drift has been studied in {\em e.g.} Stephani models \cite{drift_stephani, drift_stephani2}, Bianchi models \cite{drift_in_bianchi}, models of holographic dark energy \cite{holo2, holo3}, $f(R)$ models \cite{drift_DE_and_F(R)}, models with a varying value of $c$ \cite{drift_with_varying_c}, timescape models \cite{timescape}, and a variety of different types of dark energy models and models of modified gravity \cite{dritf_test_interacting_DE, drift_quintessence, drift_strange_DE, drift_omegaDM_and_DE, drift_DE, drift_to_test_DE, drift_with_codex, drift_test_of_lcdm, drift_test_ldcm3, drift_test_of_lcdm2, standard, MG,4th_order}. The studies show that different models will generally lead to different predictions of $\delta z$, even in cases where {\em e.g.} redshift-distance relations are similar.
\newline\newline
There is currently much focus on how inhomogeneities can affect observations and their interpretation. One particularly intriguing effect of inhomogeneities is cosmic backreaction which is a consequence of the noncommutativity of temporal development and spatial averaging in general relativity (see {\em e.g.} \cite{bc_review, bc_fluidI, bc_fluidII, also_2region} for introductions and reviews to cosmic backreaction). This effect leads the average evolution of inhomogeneous spacetimes to generally deviate from that of an FLRW spacetime even if the inhomogeneous spacetime has a homogeneity scale. Cosmic backreaction can lead to on-average accelerated expansion of the universe even though the local expansion is decelerating everywhere (see {\em e.g.} \cite{average_LTB, average_LTB2, average_LTB3,Syksy_2region, Syksy_2region2} for explicit examples and {\em e.g.} \cite{talk, incl_Newtonian, status_report} for general considerations). This has led to the backreaction conjecture which proposes that the seeming need for dark energy is simply the result of neglecting cosmic backreaction when interpreting observations through FLRW predictions.
\newline\indent
Cosmic backreaction can in principle mimic a cosmological constant \cite{bc_as_lamdba_unlikely}, although studies indicate that this is unlikely at least within the restricted setting of perturbation theory \cite{bc_mimic_lambda_but_generally_extra_terms}. In general, the backreaction conjecture is expected to lead to a different expansion history than that predicted by the standard model. As a result, redshift drift measurements in agreement with standard model predictions could be considered basis for rejecting the backreaction conjecture from the pool of viable physical explanations of dark energy. The purpose of this work is to assess the validity of such a statement. In order to do this, an expression for the average redshift drift in a statistically homogeneous and isotropic dust cosmology is derived. The resulting expression is valid even in the case of significant backreaction. By using two-region models (see section \ref{sec:2region}), the expression is used to study the possibility of obtaining average redshift drifts in models without dark energy, that are observationally indistinguishable from the redshift drift of the standard model.

\section{Redshift drift in a statistically homogeneous and isotropic dust universe}\label{sec:drift}
In \cite{Syksy_average} (see also \cite{Syksy_average2}), the expression for the redshift in a (spatially) statistically homogeneous and isotropic dust universe was derived. That derivation is the basis for obtaining an expression for the redshift drift in a statistically homogeneous and isotropic dust universe. It will thus be summarized below, functioning as a means of introducing the physical setup and notational conventions.
\newline\newline
It is assumed that the dust is irrotational, implying that the spacetime can be foliated with spatial hypersurfaces orthogonal to the velocity field of the dust \cite{golden_oldie}, {\em i.e.} the dust is comoving and the metric can be assumed to take a form corresponding to the line element:
\begin{equation}
ds^2 = -c^2dt^2 + g_{ij}dx^idx^j
\end{equation}
The components of the spatial metric $g_{ij}$ are arbitrary except that the metric is assumed to describe a statistically homogeneous and isotropic space.
\newline\indent
The velocity field of an observer comoving with the dust is denoted by $u^{\mu}$ and can be split into an expansion part, a shear part and a vorticity part as given by the equation (see {\em e.g.} \cite{golden_oldie, god_bog} or chapter 9 of \cite{apple_bog}):
\begin{equation}
u_{\mu;\nu} = \frac{1}{3}\theta h_{\mu\nu} + \sigma_{\mu\nu} + \omega_{\mu\nu}
\end{equation}
A subscripted semicolon denotes covariant differentiation. The tensor $h_{\mu\nu}$ is the projection tensor projecting onto the 3-space orthogonal to $u^{\mu}$. In the case of irrotational dust, the vorticity vanishes and so $\omega_{\mu\nu}$ is assumed to be zero in the following. It is then the expansion scalar $\theta :=u^{\mu}_{;\mu}$ and the traceless shear tensor
\footnote{The standard notation $T_{(\mu\nu)}:=\frac{1}{2}\left( T_{\mu\nu} + T_{\nu\mu}\right)$ is used.}
$\sigma_{\mu\nu}:=u_{(\mu;\nu)}-\frac{1}{3}\theta h_{\mu\nu}$ which will be important.
\newline\indent
Using the setup introduced above, an analysis in \cite{Syksy_average} shows that the redshift in a dust spacetime can be written as an integral along the given null-geodesic as follows:
\begin{equation}\label{eq:redshift}
1+z = \exp\left( \int_{t_e}^{t_0}dt \left[ \frac{1}{3}\theta + \sigma_{\mu\nu}e^{\mu}e^{\nu}\right] \right) 
\end{equation}
$e^{\mu}$ is a unit vector field orthogonal to $u^{\mu}$ and gives the propagation direction of the light ray as seen by an observer moving with velocity $u^{\mu}$, {\em i.e.} $k^{\mu} = -u^{\alpha}k_{\alpha}\left(u^{\mu} + e^{\mu} \right) $.
\newline\indent
This expression is in \cite{Syksy_average} related to average quantities, where the average is defined following the Buchert averaging scheme (see {\em e.g.} \cite{bc_fluidI,bc_review}). Using this scheme, the spatial average of a scalar field $\Psi$ over a comoving spatial domain $D$ is defined as:
\begin{equation}
\left\langle \Psi\right\rangle  :=\frac{1}{V_D}\int_{D}^{}dV\Psi = \frac{\int_{D}^{}dV\Psi}{\int_{D}^{}dV}
\end{equation}
The spatial volume element is given by $dV = \sqrt{|\det(g_{ij})|}d^3x$.
\newline\indent
Using this definition of scalar averages, a scale factor can be introduced as $\left\langle a\right\rangle  :=\left(\frac{V_D}{V_{D_0}} \right) ^{1/3}$.
\newline\newline
The expression for the redshift given in equation (\ref{eq:redshift}) is in \cite{Syksy_average} related to the average redshift by arguments that will only be summarized here. The reader is referred to \cite{Syksy_average, Syksy_average2} for further details.
\newline\indent
In a statistically homogeneous and isotropic universe, there is no preferred direction and so the projection of the shear along $e^{\mu}$ should be small when integrated over distances that are long compared to the homogeneity scale. The integral will only vanish identically in an exactly homogeneous and isotropic universe, but in a statistically homogeneous and isotropic universe the term $\sigma_{\mu\nu}e^{\mu}e^{\nu}$ will lead to a subdominant contribution in the integral above, compared to the term $\frac{1}{3}\theta$.
\newline\indent
The expansion scalar can be split into an average part and a fluctuation field $\Delta \theta$ such that $\theta = \left\langle \theta \right\rangle + \Delta \theta $. $\Delta \theta$ is not necessarily small compared to $\left\langle \theta \right\rangle $, but if structures evolve slowly compared to the time it takes light to travel the distance of the homogeneity scale, then the contribution of $\Delta \theta$ to the integral above should be much smaller than that of  $\left\langle \theta \right\rangle $; if structures are approximately static during the time it takes a light ray to travel a distance corresponding to the homogeneity scale, then the redshift gained after traveling such a distance should be the same for all geodesics (up to statistical fluctuations). Thus, $\Delta \theta$ should only lead to statistical fluctuations of measured redshifts. A rough estimate of the contribution of $\Delta \theta$ to the above integral is given in \cite{Syksy_average} with the result that the contribution can be expected to be of the order $10^{-3}$.
\newline\indent
One point of concern with the above discussion is that it is tacitly assumed that light rays ``sample" space ``fairly" when they propagate through spacetime, {\em i.e.} it is assumed that light rays do not consistently avoid or prefer a specific type of region such as voids or large overdensities. If this assumption is incorrect, the integral of $\Delta\theta$ along a null-ray will be equal to a non-zero off-set, instead of being equal to zero, up to statistical fluctuations. This will again lead the actual average observed redshift to deviate from the expression in equation \ref{eq:z_av} below. It was for instance found in \cite{Sofie2} that if voids have a narrow shape and are highly anisotropic, light rays will consistently avoid the central part of the voids. However, a study of averaging over several light rays in swiss cheese models based on more realistic void profiles indicates that light rays on average will sample space fairly \cite{Dallas_cheese}. An unfair sampling of space could also be expected to occur because baryonic matter bound in high-density structures is impermeable to light and so any light ray that reaches such highly dense region will never reach us. Since most matter is dark, it seems likely that this should not lead to a significant bias of the average observed redshift.
\newline\indent
Overall, it seems reasonable to assume that observed light rays will sample space fairly, although a more thorough study of this may be appropriate
\footnote{It is noteworthy that the result of \cite{Syksy_average,Syksy_average2}, indicating that the Dyer-Roeder approximation for distance-redshift relations is incorrect, may change if observed light rays are not assumed to sample spacetime fairly. An unfair sampling of space by light rays is exactly the assumption upon which the Dyer-Roeder expression is based.}
.
\newline\newline
From the above discussion, it follows that the redshift will approximately be equal to its average and can be written as:
\begin{equation}\label{eq:z_av}
1+z \approx 1+\left\langle z\right\rangle  = \exp\left( \int_{t_e}^{t_0}dt \frac{1}{3}\left\langle \theta\right\rangle  \right) 
\end{equation}
The average redshift can be understood as the average redshift of light emitted from a given spatial hypersurface and observed at $t = t_0$ by an observer comoving with the dust.
\newline\indent
It follows straightforward from the Buchert averaging scheme that $3\frac{\left\langle a\right\rangle _{,t}}{\left\langle a\right\rangle } = \left\langle \theta\right\rangle $, where subscripted commas followed by one or more coordinates indicate partial differentiation with respect to the indicated coordinate(s). Thus, the above expression reduces to:
\begin{equation}
1+z\approx 1+\left\langle z\right\rangle  = \frac{\left\langle a(t_0)\right\rangle }{\left\langle a(t_e)\right\rangle }
\end{equation}
An example of the difference between the average redshift and the exact redshift along individual rays can be seen in figure 9 of \cite{Tardis} where swiss-cheese models based on LTB models with significant backreaction are studied.

\subsection{The average redshift drift}
The redshift drift measured in a time interval $\delta t_0$ is to first order given by $\delta z =\frac{dz}{dt_0} \delta t_0= z(t_e+\delta t_e) - z(t_e)$. According to the previous discussion this can be written as:
\begin{equation}
\begin{split}
\delta z = \exp\left( \int_{t_e+\delta t_e}^{t_0 + \delta t_0}dt \left[ \frac{1}{3}\theta + \sigma_{\mu\nu}e^{\mu}e^{\nu}\right] \right) - \exp\left( \int_{t_e}^{t_0}dt \left[ \frac{1}{3}\theta + \sigma_{\mu\nu}e^{\mu}e^{\nu}\right] \right)
\\ 
\approx \exp\left( \int_{t_e + \delta t_e}^{t_0 + \delta t_0}dt \frac{1}{3}\left\langle \theta\right\rangle  \right) - \exp\left( \int_{t_e}^{t_0}dt\frac{1}{3}\left\langle \theta\right\rangle \right) = \frac{\left\langle a(t_0 + \delta t_0)\right\rangle }{\left\langle a(t_e + \delta t_e)\right\rangle } - \frac{\left\langle a(t_0)\right\rangle }{\left\langle a(t_e)\right\rangle }
\end{split}
\end{equation}
Aside from the brackets indicating spatial averaging, this expression is identical to the expression for the redshift drift in FLRW models. It can be written in a more convenient form by using the Taylor expansion $\left\langle a(t+\delta t)\right\rangle \approx \left\langle a(t)\right\rangle  + \left\langle a(t)\right\rangle _{,t}\delta t$. Inserting this into the above equation yields:
\begin{equation}
\delta z  \approx (1+\left\langle z\right\rangle )\left\langle H_0\right\rangle \delta t_0 - (1+\left\langle z\right\rangle )\left\langle H(t_e)\right\rangle \delta t_e
\end{equation}
The Hubble parameter has here been defined by $\left\langle H\right\rangle := \frac{\left\langle a\right\rangle _{,t}}{\left\langle a\right\rangle }$.
\newline\indent
The last step is to obtain a relation between $\delta t_e$ and $\delta t_0$. Since there are no peculiar velocities with the assumed foliation of spacetime, $\frac{\delta t_0}{\delta t_e} = 1+z$ follows directly from the geometric optics approximation (see {\em e.g.} chapters 7 and 15 of \cite{god_bog} or section 3.2 of \cite{lens_bog}). As illustrated by the equality $\frac{\delta t_0}{\delta t_e} = 1+z$, $\delta t_e$ is a quantity that in principle needs to be averaged. However, $z\approx \left\langle z\right\rangle $ implies that $\delta t_e \approx \left\langle\delta t_e \right\rangle $, where $\left\langle \delta t_e\right\rangle $ is the time interval corresponding to $\delta t_0$ averaged over the spatial hypersurface at $t = t_e$.
\newline\indent
The above considerations show that the expression for the redshift drift can be written as\footnote{Please see arXiv:1909.13489 for a correction.}:
\begin{equation}\label{eq:drift_av}
\delta z \approx \left\langle \delta z \right\rangle =  \left[ (1+\left\langle z\right\rangle )\left\langle H_0\right\rangle  - \left\langle H(t_e)\right\rangle \right] \delta t_0
\end{equation}
Aside from the indicated approximation and the brackets indicating spatial averaging, this is exactly the same expression as for FLRW spacetimes. The expression is also the same as that obtained for timescape backreaction models \cite{timescape}, where the result is obtained by introducing a local, average metric based on``dressed" parameters (see {\em e.g.} \cite{dressed} for an introduction to the concept of dressed parameters). The expression for the average redshift drift given in equation (\ref{eq:drift_av}) is in principle also stated in \cite{Syksy_average2}. In relation to that statement, concern is expressed regarding the potential importance of fluctuations around $\left\langle \delta z\right\rangle $ due to the small size of $\delta z$ itself. This point is discussed in subsection \ref{sec:caveats}, where the justification of equation (\ref{eq:drift_av}) is studied more closely.
\newline\newline
The expression in equation (\ref{eq:drift_av}) is valid not only for dust models, but also models containing one or more dark energy components with vanishing sound speed; such components will not have pressure gradients and will thus follow a geodesic flow. If the individual components are comoving initially, they will stay comoving and so the comoving, synchronous foliation of spacetime with space orthogonal to the flow lines is permitted. 
\newline\indent
Before moving on, it should be noted that an expression for the average redshift drift has earlier been derived using light-cone averages (see \cite{drift_lightcone}). However, the spatially averaged redshift drift is related to observables; the redshift drift of a luminous object deviates from the spatially averaged redshift drift only by statistical fluctuations. Although the light-cone approach certainly has its merits, it does thus not seem to be immediately necessary to introduce light-cone averaging for redshift drift considerations.

\subsection{Deviations from the average redshift drift}\label{sec:caveats}
Equation (\ref{eq:drift_av}) gives an expression for the average redshift drift measured by an observer comoving with the dust in a statistically homogeneous and isotropic universe. In that equation, it is also implied that deviations between the average redshift drift and the exact redshift drift of a given luminous object are small. That assessment is considered more carefully in this subsection.
\newline\newline
The exact redshift drift can be written as:
\begin{equation}
\begin{split}
\delta z =  X_{s1}\times\frac{\left\langle a(t_0 + \delta t_0)\right\rangle }{\left\langle a(t_e + \delta t_e)\right\rangle } - X_{s2}\times\frac{\left\langle a(t_0)\right\rangle }{\left\langle a(t_e)\right\rangle }
\end{split}
\end{equation}
The factors  $X_{s1}$ and $X_{s2}$ depend on the particular source and observer and are given by:
\begin{equation}
\begin{split}
X_{s1}: = \exp\left( \int_{t_e+\delta t_e}^{t_0 + \delta t_0}dt \left[ \frac{1}{3}\Delta \theta + \sigma_{\mu\nu}e^{\mu}e^{\nu}\right] \right)
\\
X_{s2} := \exp\left( \int_{t_e}^{t_0}dt \left[ \frac{1}{3}\Delta \theta + \sigma_{\mu\nu}e^{\mu}e^{\nu}\right] \right)
\end{split}
\end{equation}
According to the arguments given in \cite{Syksy_average} and summarized in the previous subsection, each of these two factors are approximately equal to one. In addition, the two factors are approximately equal under the assumption that $\delta t_0$ and $\delta t_e$ are small compared to both the scale of cosmic structure formation and compared to the time interval $t_0-t_e$. These two assumptions are already inherent for discussing the redshift drift in the range $2\lesssim z \lesssim 5$, and stating that $X_{s1}\approx X_{s2}$ is similar to noting that the redshift drift is a small quantity (typically of order $10^{-10}-10^{-8}$ on a time scale of $\sim $ 10 years for a source at $z \approx 4$). The former part of the assumption ensures that the light emitted from a given source and observed at the two different epochs has traveled along only negligibly different spacetime paths. The assumption then ensures that the functions under the two integrals can be considered the same even though they are in principle traced along two different  null-geodesics. The latter assumption is necessary for the obvious numerical reason that two integrals over the same integrand are only equal if the integration limits are the same (or if the integrand vanishes or is self-canceling where the integration limits differ which is also possible in the considered physical setup).
\newline\indent
Since $X_{s1}\approx X_{s2}\approx 1$ and $\delta t_e \approx \left\langle \delta t_e \right\rangle $, the approximation $\delta z \approx \left\langle \delta z\right\rangle $ should be fairly safe, especially considering that real observations will presumably be made with several sources at similar redshifts such that an average can be made to reduce errors due to statistical fluctuations.
\newline\newline
The question of whether deviations from the average value are important for redshift drift observations has also been considered elsewhere with the focus mainly being on peculiar motion.  Peculiar motion was neglected above since the spacetime foliation was chosen to be comoving. In the real universe, peculiar motion may be an important source of contamination of redshift drift measurements. This possibility has been studied elsewhere with the conclusions generally being that peculiar velocities will have negligible effect on redshift drift observations while peculiar acceleration might have a significant effect which should however vanish upon averaging over several objects \cite{Loeb, perturbation_drift, drift_pert2,Early, perturbation_drift, drift_observational_feasibility,sigma, clouds} (see {\em e.g.} also \cite{ peculiar_acc}). In appendix \ref{app:cheese_and_voids}, the exact redshift drift is computed in swiss-cheese models containing dust and dark energy. These computations are in accordance with the assumption that fluctuations of the redshift drift about the average value are small.
\newline\indent
The local environment of the observer may bias observations and this is not taken into account in the expression given in equation (\ref{eq:drift_av}). However, it is illustrated in appendix \ref{app:cheese_and_voids} that very large and deep voids are needed in order to compromise redshift drift measurements. Effects of a local overdensity has not been studied as such are expected to be of the same size as those caused by a local underdensity.

\subsection{Redshift drift in two-region models}\label{sec:2region}
Two-region models \cite{Syksy_2region, Syksy_2region2, also_2region} (se also {\em e.g.} \cite{Syksy_peak, Syksy_peak2}) are simple toy-models that can be used to illustrate consequences of non-vanishing cosmic backreaction. Two-region models consist of two non-interacting regions that each behave according to the Friedmann equations, but where the FLRW parameters are different in the two regions.
\newline\indent
The average scale factor of a two-region model is given by $\left\langle a\right\rangle  = \left( \frac{a_u^3 + a_o^3}{a_{u,0}^3 +a_{o,0}^3 }\right) ^{1/3}$, where subscripts $o,u$ indicate the two separate regions. Since the two individual regions are non-interacting, the average Hubble parameter is simply given by:
\begin{equation}
\left\langle H\right\rangle  = H_u\frac{a_u^3}{a_u^3+a_o^3} + H_o\frac{a_o^3}{a_u^3+a_o^3}
\end{equation}
The two-region models studied here are based on the two-region model considered in {\em e.g.} \cite{Syksy_2region}. In this type of two-region models, the underdense region is completely empty while the overdense region is a closed matter-only FLRW model. In such a model, cosmic time and the scale factors can be related to the development angle $\phi$ of the overdense region as follows:
\begin{equation}
\begin{split}
t = t_0\frac{\phi-\sin(\phi)}{\phi_0-\sin(\phi_0)}\\
a_u = \frac{f_u^{1/3}}{\pi}\left( \phi - \sin(\phi)\right)\\
a_o = \frac{f_o^{1/3}}{2} \left( 1-\cos(\phi)\right) 
\end{split}
\end{equation}
The models studied here are defined by setting $\phi_0 = 3/2\pi$ and by requiring that the average Hubble constant takes the value $\left\langle H_0\right\rangle  = 70$km/s/Mpc. $f_o$ and $f_u = 1-f_o$ are the volume fractions of the over- and underdense regions at $\phi = \pi$. The parameter $f_o$ is varied in order to obtain two-region models with different expansion histories.
\newline\newline
Two-region models can be related to statistically homogeneous and isotropic universes as follows. The spatial hypersurfaces of a statistically homogeneous and isotropic universe can be divided into areas that are overdense and areas that are underdense compared to some threshold value. The two-region models can be considered the simplest attempt to include effects of inhomogeneities by splitting the spacetime into one part describing the total underdense portion of the spacetime and one part describing the entire overdense portion. Both regions are approximated as being described by an FLRW model. In the particular two-region models used here, the underdense region is for simplicity modeled as empty, but it could instead have been modeled {\em e.g.} as an open matter-only FLRW region (see {\em e.g.} \cite{Syksy_boehm} for an example of such a model). When viewed in this manner, the two-region models can be considered a simple subclass of the multiscale models introduced in \cite{multiscale} (see {\em e.g.} also \cite{multiscale2})\footnote{This was pointed out to us by the anonymous referee.}.
\newline\indent
Cosmic backreaction is an effect that occurs because of spatial averaging. Clearly, much of this averaging is neglected in the two-region models if they are to be considered as just described. The two-region models are nonetheless useful for proof-of-principle studies of effects of cosmic backreaction.

\section{The spectral velocity shift in particular two-region models compared to expected $\Lambda$CDM observations}\label{sec:results}
Redshift drift leads to a spectral velocity shift $\delta v :=\frac{c\delta z}{1+z}$. The accuracy of a $\delta v$ measurement of high-redshift QSO spectra with the next generation of ELTs was investigated in \cite{sigma}, leading to the following expression for the expected measurement uncertainties:
\begin{equation}\label{eq:sigma}
\sigma_{\delta v} = \alpha\left(\frac{S/N}{2370} \right)^{-1}\left( \frac{N_{QSO}}{30}\right) ^{-1/2}\left( \frac{1+z_{QSO}}{5}\right) ^\beta cm/s
\end{equation}
$S/N$ is the signal-to-noise ratio per $0.0125\text{\AA}$ pixel, $N_{QSO}$ is the number of targets at a given redshift $z_{QSO}$, and $\beta = -1.7$ for $z_{QSO}\leq4$ while $\beta=-0.9$ otherwise. The factor $\alpha$ is equal to 2 if only $HI$ absorption lines are considered, and reduces to $1.35$ if all available absorption lines are included. Following \cite{drift_test_of_lcdm2}, the values $N_{QSO} = 40$, $S/N = 3000$ and $\delta t_0 = 30$ years will be used in the following when estimating uncertainties in $\delta v$ (unless otherwise stated). The 40 observed spectra will be assumed equally distributed over 5 redshift bins with equally spaced centers in the range $2\leq z \leq 5$. The smaller factor $\alpha = 1.35$ will be used.
\newline\indent
The results shown below include error bars based on the specifications given above assuming that the true cosmological model is given by a flat $\Lambda$CDM model with $\Omega_{m,0} = 0.3$ and $H_0 = 70$km/s/Mpc. Models with redshift drifts falling inside the error bars would be observationally indistinguishable from this $\Lambda$CDM model as far as redshift drift measurements are concerned. In other words, models with their redshift drift falling inside the error bars would be mistaken for the $\Lambda$CDM model when redshift drift observations are interpreted using FLRW cosmology.
\newline\newline

\begin{figure}[htb!]
\centering
\subfigure[]{
\includegraphics[scale = 0.8]{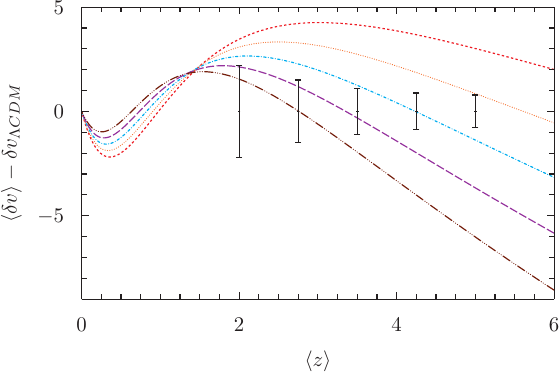}
}
\subfigure[]{
\includegraphics[scale = 0.8]{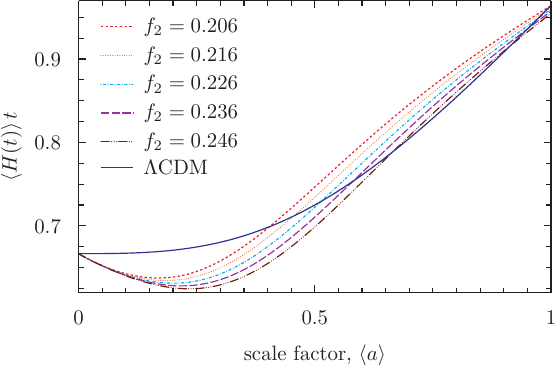}
}
\caption{Velocity shift and $\left\langle H\right\rangle t$ for the two-region models and a flat $\Lambda$CDM model specified by $\Omega_{m,0} = 0.3$ and $H_0 = 70$km/s/Mpc. The velocity shift is shown as a deviation from that of the $\Lambda$CDM model. The velocity shift is given in units of cm/s. Error bars represent the expected uncertainty of ELT redshift drift measurements.}
\label{fig:2region}
\end{figure}
\begin{figure}[htb!]
\centering
\subfigure[]{
\includegraphics[scale = 0.8]{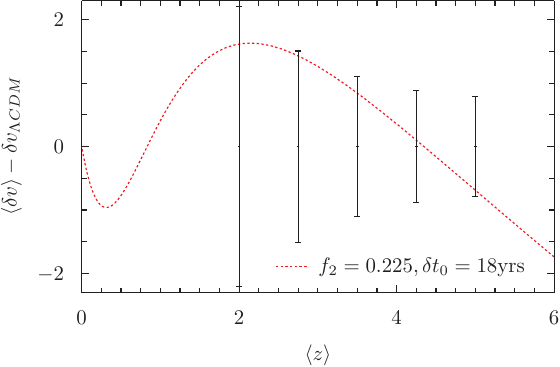}
}
\subfigure[]{
\includegraphics[scale = 0.8]{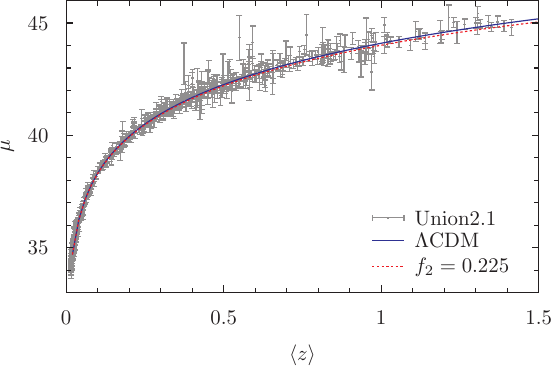}
}
\caption{Velocity shift and angular diameter distance of a two-region model and the flat $\Lambda$CDM model specified by $\Omega_{m,0} = 0.3$ and $H_0=70$km/s/Mpc. The velocity shift is shown as a deviation from that of the $\Lambda$CDM model. The velocity shift is given in units of cm/s. The redshift-distance relation is plotted together with Union2.1 data \cite{union2}. The Union 2.1 data is available at http://supernova.lbl.gov/union/.}
\label{fig:18years}
\end{figure}

The difference between the velocity shift of a $\Lambda$CDM model and a variety of (similar) two-region models is shown in figure \ref{fig:2region}. The average Hubble parameters of the models are also shown, illustrating the differences in the expansion histories of the models. As seen, all of the two-region models' velocity shifts deviate from the $\Lambda$CDM model's at a level above the accuracy expected for the ELT measurements. This indicates that measurements of $\delta z$ in the time interval $\delta t_0 = 30$ years would be able to distinguish between the two-region toy-models and the $\Lambda$CDM model. However, it is not unrealistic to expect that $\delta z$ measurements will be made with $\delta t_0<30$ years. For instance, several of the papers cited earlier use $\delta t_0 = 10-20$ years. The measurement of a redshift at the required precision is quite time-consuming. The number of years needed in order to make sufficiently accurate measurements of the redshift drift in a time interval $\delta t_0$ will in practice most likely be several of decades longer than $\delta t_0$; a sufficiently accurate measurement of the redshift at just a single epoch is expected to require in the order of 10 years \cite{drift_test_of_lcdm2, realtime} {\em e.g} in order to achieve a high level of signal-to-noise. $\delta t_0$ is the time between two separate measurements and does not include these decade-long measuring times. Thus, curiosity induced impatience could easily lead to early redshift drift measurements based on $\delta t_0 <30$ years. Small $\delta t_0$ may also be used simply because it has been noted \cite{sigma} that measuring at more than two epochs may be beneficial for the accuracy of redshift drift measurements. One could then expect that several measurements of the redshift of specific QSOs will be made successively, decades apart over several decades, perhaps centuries, to come. In conclusion, it is interesting to see how small $\delta t_0$ must be in order for the two-region models studied here to have velocity shifts that are observationally indistinguishable from that of the $\Lambda$CDM model. The result of such a study is $\delta t_0\lesssim 18$ years. The velocity shift corresponding to $\delta t_0 = 18$ years of a particular two-region model is shown in figure \ref{fig:18years}. The figure also shows the redshift-distance relation for the particular two-region model, illustrating that the model also makes a reasonable reproduction of the observed redshift-distance relation of supernovae. The redshift-distance relation for the two-region model has been computed following the results of \cite{Syksy_average, Syksy_average2} (see also \cite{Syksy_boehm}) which include an ODE for the average angular diameter distance, $D_A$, in terms of the average redshift:
\begin{equation}
\frac{d^2D_A}{d\left\langle z\right\rangle ^2} = -\frac{4\pi G_N \left\langle \rho\right\rangle }{\left(\left\langle H\right\rangle (1+\left\langle z\right\rangle ) \right) ^2} D_A - \frac{dD_A}{d\left\langle z\right\rangle }\left( \frac{2}{1+\left\langle z\right\rangle } + \frac{1}{\left\langle H\right\rangle }\frac{d\left\langle H\right\rangle }{d\left\langle z\right\rangle }\right) 
\end{equation}
The initial conditions are given by $D_A(0) = 0, \frac{dD_A}{d\left\langle z\right\rangle }|_{\left\langle z\right\rangle =0} = \frac{c}{\left\langle H_0\right\rangle }$.
\newline\indent

\begin{figure}[htb!]
\centering
\subfigure[]{
\includegraphics[scale = 1]{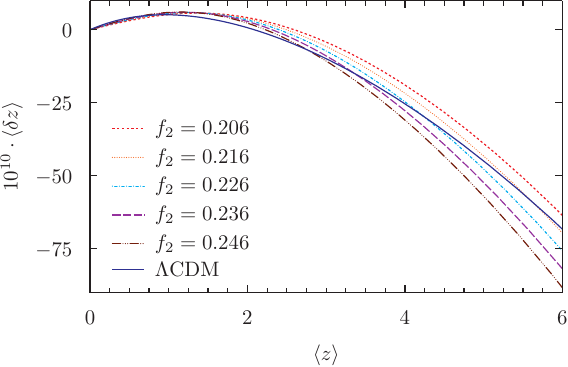}
}
\caption{Redshift drift of two-region models compared to the redshift drift of the flat $\Lambda$CDM model specified by $\Omega_{m,0} = 0.3$ and $H_0 = 70$km/s/Mpc.}
\label{fig:2region_zdrift}
\end{figure}

Note that it is possible to construct two-region models with a redshift drift that is observationally indistinguishable from that of the $\Lambda$CDM model even for $\delta t_0 \geq 30$ years. The caveat is that this requires abandoning the restriction that the average Hubble parameter has a present time value equal to the Hubble constant of the $\Lambda$CDM model used in the comparisons ($H_0 = $70km/s/Mpc). Such models will generally have a redshift-distance relation that deviates significantly from that of the $\Lambda$CDM model.
\newline\newline
Lastly it is noted that, as shown in figure \ref{fig:2region_zdrift},  the redshift drift of all the studied two-region models is positive at low redshifts. This illustrates the fact that observing a positive redshift drift at low redshifts is {\em not} a cause to reject any cosmological model that does not include a locally accelerating expansion of the Universe
\footnote{Positive redshift drifts in models without dark energy have also been reported elsewhere, for instance in Stephani models \cite{drift_stephani, drift_stephani2} and in timescape backreaction models \cite{timescape}.}
. One caveat with this statement is the issue of the relevance of averaging at low redshifts which corresponds to averaging over small volumes. As discussed in the next section, this is not an issue relevant for redshift drift considerations as the redshift drift is primarily expected to be measured at redshifts above $z\approx 2$ with the redshift drift at smaller redshifts being inferable from high-redshift observations.

\section{Discussion}
An expression for the average redshift drift in a statistically homogeneous and isotropic universe was derived. The expression has the same form as for FLRW models, but the quantities that enter are now average quantities which because of cosmic backreaction may behave differently than those of FLRW models. The redshift drift of an individual object will in general deviate from the average value but such statistical fluctuations are estimated to be small. This estimate is based partially on the results of others and partially on a study based on swiss-cheese models presented in appendix \ref{app:cheese_and_voids}.
\newline\indent
The average expression for the redshift drift is only useful for observations above the homogeneity scale. This could be expected to be a problem since the redshift drift at low redshifts is particularly interesting; the redshift drift at low redshifts is positive for the $\Lambda$CDM model while it is negative for many models without a dark energy component. However, since redshift drift measurements with the ELT are expected to be made primarily in the redshift range $2\lesssim z\lesssim5$, this issue does not pose an immediate problem for redshift drift measurements.  In {\em e.g.} \cite{sigma} it is explained how high-redshift measurements of the redshift drift can lead to information about the redshift drift at low redshifts.
\newline\indent
In order to study the possibility of using redshift drift measurements to observationally distinguish between the standard model and the backreaction conjecture, the average redshift drift in different two-region models has been computed. It was possible to construct simple two-region models that lead to a velocity shift indistinguishable from that of the $\Lambda$CDM model in cases with $\delta t_0 \lesssim 18$ years. For longer $\delta t_0$, the differences between the predictions of the two-region models and the $\Lambda$CDM model become larger than the expected uncertainty in velocity shift measurements with the next generation of extremely large telescopes. Exceptions to this statement can be found if the redshift-distance relations of the two-region models are permitted to deviate significantly from that of the $\Lambda$CDM model. 
\newline\indent
The two-region models are simple (toy-)models of inhomogeneous universes and so a quantitative analysis based on these models cannot be used to conclude with certainty to what extent the backreaction conjecture and standard model can be distinguished through redshift drift measurements. The results obtained here instead simply illustrate that backreaction {\em can} lead to an expansion history of the Universe which is indistinguishable from that of the standard model with redshift drift measurements, at least if $\delta t_0 \lesssim 18$ years. Taking into account that the measurement of a redshift value at a single epoch is expected to take around 10 years, it will thus be long (several decades to a century) before redshift drift measurements become precise enough for them to seriously distinguish between standard cosmological models and models based on the backreaction conjecture. In addition, the important feature that the redshift drift of the $\Lambda$CDM model is positive at low redshifts is also fulfilled for many two-region models.

\section{Acknowledgments}
We thank Chris Clarkson for a discussion regarding the work presented in \cite{Copernican_drift}, and the anonymous referee for suggestions for improving the manuscript.
\newline\indent
The numerical codes written for this project utilize the GNU Scientific Library\footnote{http://www.gnu.org/}.

\appendix
\section{Redshift drift in swiss-cheese and local void models}\label{app:cheese_and_voids}
In this appendix, spherically symmetric void models and swiss-cheese models based on these will be used to study inhomogeneity induced deviations of the redshift drift from its average value. The appendix is divided into four subsections. First, the spherically symmetric models are introduced. After that, the procedure for computing the redshift drift in these models is summarized, and finally the obtained results are shown.

\subsection{Spherically symmetric models with inhomogeneous dust and dark energy}
The exact solution to Einstein's field equations for spherically symmetric cosmological models containing $n$ decoupled, non-comoving perfect fluids was presented by  Valerio Marra and Mikko Paakkonen in \cite{Mikko}. A special limit of these are the spherically symmetric models with an inhomogeneous dust (dark matter) component and an inhomogeneous dark energy component. These models can be simplified by restricting the dark energy sound speed be zero such that the dust and dark energy is comoving (assuming that they start out comoving). Such simplified models are the ones that will be considered in the following, and will for notational simplicity be referred to as the ``MP'' models. 
\newline\indent
Below, a summary of the dynamics of MP models is given, followed by a specification of the particular models used here. The reader is referred to \cite{Mikko, Mikko2} for details.
\newline\newline
The line element of the MP models can be written exactly as the line element of the LTB models {\em i.e.}:
\begin{equation}
ds^2 = -c^2dt^2 + \frac{A_{,r}(t,r)^2}{1-k(r)}dr^2 + A(t,r)^2d\Omega^2
\end{equation}
The dynamical equations can be written as:
\begin{equation}
\begin{split}
A_{,t}^2 = c^2\left( \frac{2M}{A} - k\right) \\
A_{,tr} = c^2\left( \frac{M_{,r}}{AA_{,t}} - \frac{MA_{,r}}{A^2A_{,t}} - \frac{k_{,r}}{2A_{,t}} \right)\\
M_{,t} =-\frac{4\pi G_N}{c^4}A^2A_{,t}p_{de} \\
(\rho_{de})_{,t} = -\left(\rho_{de} + \frac{p_{de}}{c^2} \right)\left(\frac{A_{,tr}}{A_{,r}} + 2\frac{A_{,t}}{A} \right)  
\end{split}
\end{equation}
The topmost equation is a rewriting of the equation defining $M = M(t,r)$. The main difference between the above set of equations and the equivalent equations for LTB models is that $M$ has a time dependence due to the pressure of the dark energy component.
\newline\indent
The above four equations make up a set of coupled ODEs that can be solved simultaneously. The initial conditions which have been used here will be described below. There is no need for a differential equation for the dark matter component's density field as this can be obtained simply as $\rho_{dm}(t,r) = \rho_{dm}(t_i,r)\frac{A^2(t_i,r)A_{,r}(t_i,r)}{A^2(t,r)A_{,r}(t,r)}$.
\newline\newline
The MP models will be used to construct single void models that reduce to FLRW models either outside some specific boundary value $r_b$ of the radial coordinate, or asymptotically. This FLRW model will be referred to as the background of the model.
\newline\indent
The dark energy component has an equation of state parameter $\omega_{de}<-1/3$ and will be sub-dominant to the dark matter component at early times. Hence, at early times, the dark energy component can be ignored and the model reduces to a regular LTB model. By setting the initial conditions at early times, these can thus be set by approximating the MP model as an LTB model. There are then two free functions and an extra degree of freedom due to the coordinate covariance of the radial coordinate. In addition, an equation of state parameter for the dark energy component must be specified. The initial conditions will be set at the background time of last scattering ({\em i.e.} at $a = \frac{1}{1100}$, where $a$ is the scale factor of the background model).
\newline\indent
The coordinate covariance is removed by setting $A(t_i,r)=a(t_i)r$. The big bang time is set equal to zero. Under this requirement, it can be shown \cite{strong_to_weak} that the curvature function $k(r)$ can be related to the density perturbation by the approximate relation $\delta M_{dm}(t_i,r) = \frac{3}{5}k(r)\left( \frac{c}{a(t_i)H(t_i)r}\right) ^2$, where $H$ is the Hubble parameter of the background model. In this equation, the perturbation in the density field is given through $\delta M_{dm}(t_i,r) = \frac{M_{dm}(t_i,r)}{M_{dm,bg}(t_i)}-1$ and $\rho_{dm}(t_i,r) = \frac{\left( M_{dm}(t_i,r)\right) _{,r}c^2}{4\pi G_N A^2A_{,r}}$. The dark energy density perturbation is set according to $\delta M_{de} = \delta M_{dm}\frac{1+\omega_{de}}{1-3\omega_{de}}$. Since the dark energy sound speed is set equal to zero, the dark energy pressure is homogeneous, $p_{de} = p_{de,bg}$, where $p_{de,bg}$ is the pressure of the dark energy component in the background (see \cite{Mikko, Mikko2} regarding details of the dark energy component's equation of state parameter).
\newline\indent
The final specification of the initial conditions is made by choosing the functional form of $k(r)$ and by specifying background functions including $\omega_{de,bg}$. The functions studied here have $k(r)$ given by one of the two following functional forms:
\begin{equation}\label{eq:k_comp}
k(r) = -r^2k_{\text{max}}\left( \left( \frac{r}{r_b}\right) ^n - 1\right)^2 \,\,\, \text{for}\, r\leq r_b \,\,\, \text{and} \,k(r)=0\,\,\, \text{for}\, r>r_b
\end{equation}
\begin{equation}\label{eq:k_uncomp}
k(r) = -r^2k_{\text{max}}\exp\left( -\left( \frac{r-\Delta r}{\sigma}\right) ^2 \right) 
\end{equation}
The parameters $k_{\text{max}},n, r_b,\Delta r,\sigma$ are varied. In all the models, $k_{\text{max}}$ is of the order $10^{-10}-10^{-9}$, $r_b$ and $\sigma$ are always of the same order as each other, and $\Delta r \sim 10$.

\subsection{Redshift drift along radial rays in spherically symmetric models}
A set of ODEs that can be used to obtain the redshift drift along radial rays in LTB models was given in {\em e.g.} \cite{drift_LTB} and the results were later generalized to axially symmetric Szekeres models and Szekeres models with no symmetries \cite{drift_cylinder_cheese, drift_cylinder_cheese2,drift_szekeres}. Here, the ODEs presented in \cite{drift_LTB} are sufficient and will be listed below.
\newline\newline
Radial null-geodesics can be described by the two ODEs:
\begin{equation}\label{eq:MP_geodesics}
\begin{split}
\frac{dz}{dr} = (1+z)\frac{A_{,tr}}{c\sqrt{1-k}}\\
\frac{dt}{dr} = -\frac{A_{,r}}{c\sqrt{1-k}}
\end{split}
\end{equation}
Note that the sign conventions are here chosen to fit a situation with $r_{0}<r_{e}$, but the sign conventions in the ODEs may be changed to suit the reverse situation. This is needed {\em e.g.} when working with swiss-cheese models, where it can be convenient to introduce a ``global" radial coordinate.
\newline\indent
The redshift drift is defined as $\delta z := z(t + \delta t) - z(t)$, where $\delta t$ is a small time interval corresponding to the small observer-based time interval $\delta t_0$. Inserting this into the two equations above and introducing a Taylor expansion leads to the following further ODEs when only first order terms in $\delta z$ and $\delta t$ are kept:
\begin{equation}\label{eq:zdrift_LTB}
\begin{split}
\frac{d\delta z}{dr} = \frac{A_{,tr}\delta z}{c\sqrt{1-k}} + (1+z)\frac{A_{,ttr}\delta t}{c\sqrt{1-k}}\\
\frac{d\delta t}{dr} = -\frac{A_{,tr}\delta t}{c\sqrt{1-k}}
\end{split}
\end{equation}
\newline\indent
These four ODEs can be solved simultaneously to obtain the redshift drift as a function of the redshift along radial rays in MP models. In \cite{drift_LTB}, the differential equations are transformed into equations where the differentiations are with respect to the redshift. This is done under the restriction of only considering models where the redshift is monotonic along the null-geodesics. This requirement is generally not fulfilled and in particular, some of the models studied here do not fulfill it. The set of ODEs is thus kept as above, where the differentiation is with respect to the radial coordinate.
\newline\newline
Note that an analytic expression for the redshift drift as seen by a central observer in an LTB universe was derived in \cite{Copernican_drift} (and given in equation 9 of \cite{Copernican_drift}). The definition of the redshift drift in \cite{Copernican_drift} appears to be slightly different than the definition used here, although the two definitions agree in the FLRW limit. The latter is explicitly demonstrated in \cite{Copernican_drift} while the former {\em e.g.} can be seen numerically by comparing the redshift drift defined in \cite{Copernican_drift} to the redshift drift computed with equation (\ref{eq:zdrift_LTB}), or by comparing the differential equation in (\ref{eq:zdrift_LTB}) with the equivalent equation obtained by differentiating the redshift drift corresponding to equation 9 in \cite{Copernican_drift}. At least for the models studied here, the difference between the two redshift drifts is small and does not lead to qualitative differences in the study described in this appendix.
\newline\newline
Upon averaging over the entire double-structure of the void-wall models used here, the average evolution reduces to that of the background of the particular model. Thus, the models considered in the following have vanishing backreaction and the average redshift drift in the models is simply the redshift drift of the background models.

\subsection{Redshift drift: effects of local voids}
The expression for the (average) redshift drift given in equation (\ref{eq:drift_av}) does not take possible effects of the local environment into account. By computing the redshift drift in different void models, it is here studied how large a local void is likely to need to be in order for it to bias redshift drift observations at an observationally significant level. All the results shown below were obtained by setting $\delta t_0 = 30$ years.
\newline\newline
\begin{figure}[htb!]
\centering
\subfigure[]{
\includegraphics[scale = 0.8]{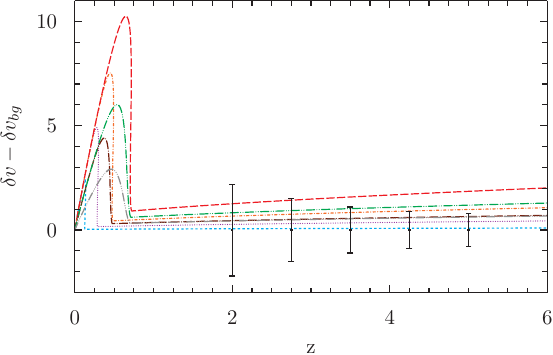}
}
\subfigure[]{
\includegraphics[scale = 0.8]{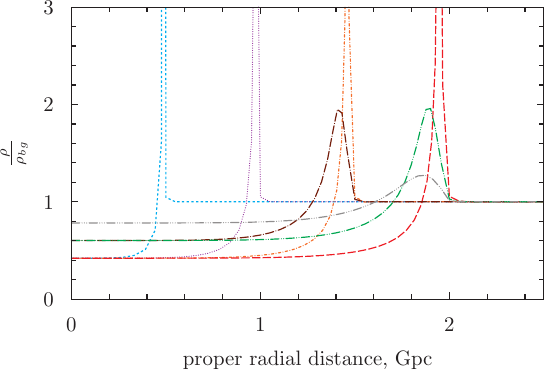}
}
\caption{The figure to the left shows the velocity shift for different void models compared to the velocity shift $\delta v_{bg}$ of the background model. The velocity shift is given in units of cm/s and the observer is placed at the origin. Error bars are included according to the prescription given in section \ref{sec:results}. The figure to the right shows the present time overdensity of the voids with the proper radial distance approximated by the metric function $A$. The density profiles that are cut at the top of the figure all reach a maximum of approximately 60. The background model is an EdS model.}
\label{fig:LTBvoids}
\end{figure}

\begin{figure}[htb!]
\centering
\subfigure[]{
\includegraphics[scale = 0.8]{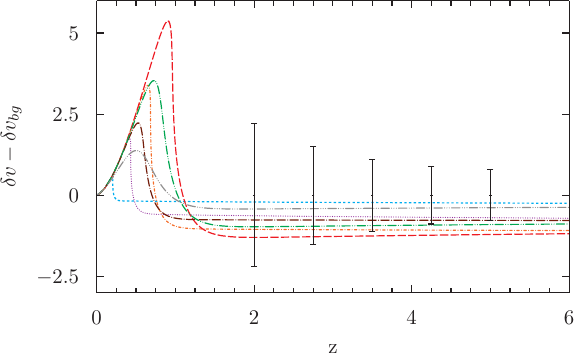}
}
\subfigure[]{
\includegraphics[scale = 0.8]{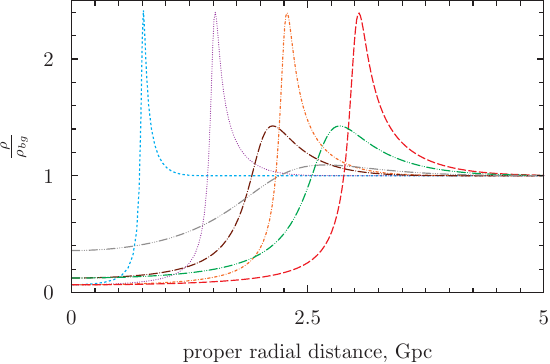}
}
\caption{The figure to the left shows the velocity shift for different void models compared to the velocity shift $\delta v_{bg}$ of the background model. The velocity shift is given in units of cm/s and the observer is placed at the origin. Error bars are included according to the prescription given in section \ref{sec:results}. The figure to the right shows the present time overdensity of the voids with the proper radial distance approximated by the metric function $A$. The background model is a flat $\Lambda$CDM model with $\Omega_{m,0} = 0.3$.}
\label{fig:LLTBvoids}
\end{figure}

\begin{figure}[htb!]
\centering
\subfigure[]{
\includegraphics[scale = 0.8]{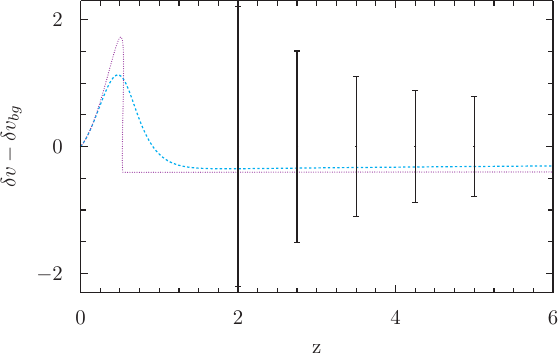}
}
\subfigure[]{
\includegraphics[scale = 0.8]{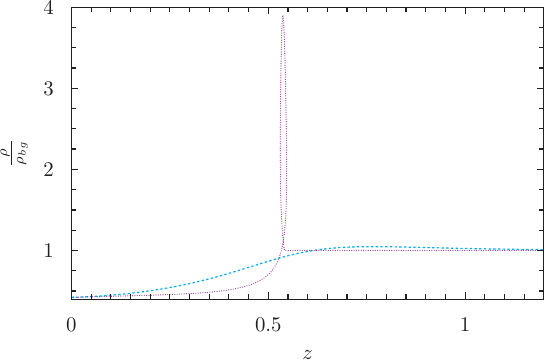}
}\par
\subfigure[]{
\includegraphics[scale = 0.8]{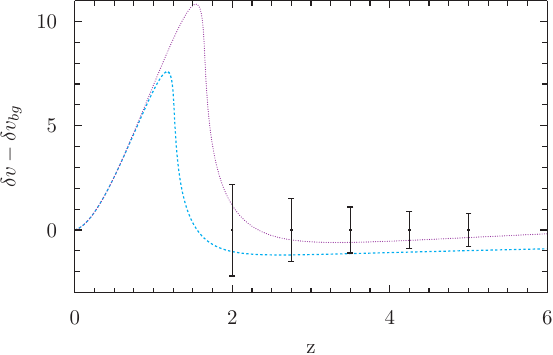}
}
\subfigure[]{
\includegraphics[scale = 0.8]{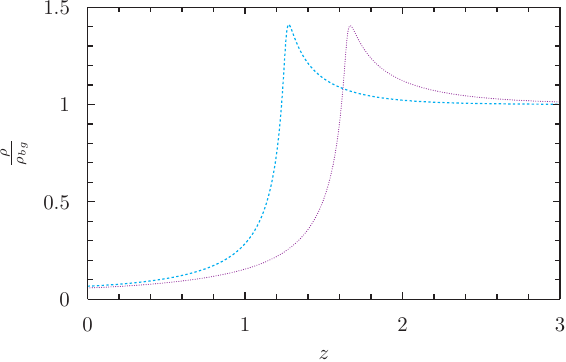}
}
\caption{Density and velocity shift as seen by an observer at the center of four different voids. The velocity shift is shown as a shift compared to the flat $\Lambda$CDM model with $\Omega_{m,0} = 0.3$ and $H_0 = 70$km/s/Mpc. In the top row, the velocity shift is compared between two voids of approximately same size and depth but with different density profiles. In the bottom row, the velocity shift is compared for two voids of different size. The velocity shift is shown in units of cm/s.}
\label{fig:compare}
\end{figure}

Figure \ref{fig:LTBvoids} shows the velocity shift in different void models with an Einstein-de Sitter (EdS) background defined by $H_0 = 70$km/s/Mpc. Figure \ref{fig:LLTBvoids} shows the same for models with a $\Lambda$CDM background defined by $H_0$ = 70km/s/Mpc, $\Omega_{dm,0} = 0.3$ and $\Omega_{\Lambda} = 0.7$. The figures indicate that voids must be very deep and large ($\sim$Gpc radius) in order for them to affect observations. Note also that the voids must be significantly larger and deeper when the background is a $\Lambda$CDM background. This is expected since the cosmological constant comprises the larger part of the present time energy density and is homogeneous. The total overdensity $\frac{\rho_{de} + \rho_{dm}}{\rho_{de,bg} + \rho_{dm,bg}}$ is thus significantly smaller than the matter overdensity $\frac{\rho_{dm}}{\rho_{dm,bg}}$ in models with the $\Lambda$CDM model as their background. As will be shown in the next subsection, the density fluctuations of dark energy components generally remain small compared to the dark matter density fluctuations of the same model. Changing $\omega_{de}$ slightly from -1 will therefore not change the result that in models dominated by dark energy, the dark matter perturbations must be larger and deeper than in pure dark matter models, in order for the density fluctuations to have significant impact on redshift drift observations.
\newline\newline
Aside from the backgrounds used in the two figures being different, the void profiles have also been modeled differently; the models with the LTB backgrounds have been modeled using $k(r)$ as in equation (\ref{eq:k_comp}) while the models with the $\Lambda$CDM background have been modeled with $k(r)$ as given in equation (\ref{eq:k_uncomp}). The void profile does not seem to be particularly important for the redshift drift. This is illustrated in the top row of figure \ref{fig:compare}. The figure shows that the void profile is mildly important for the effects on the velocity shift, but the overall effect of voids on the redshift drift/velocity shift does not seem to depend much on the particular density profile (consistent with remarks made in \cite{drift_ltb_vs_DE}). Instead, the important void characteristics are the void depth and radial extent.
\newline\newline

The results obtained above are based on studying particular inhomogeneous solutions to Einstein's equations. Other void models may in principle lead to different results. It does, however, seem reasonable to expect that the results obtained here will be valid in general at a qualitative level. {\em I.e.} it will presumably generally be the case that large voids are needed in order for local voids to affect redshift drift measurements at a significant level. Even in this case, the results here do not guarantee that large voids will be detectable with redshift drift measurements and that small voids will not affect observations. The former is illustrated in the bottom row of figure \ref{fig:compare} which shows the shift in the velocity shift compared to the background ($\Lambda$CDM) model. The velocity shift is shown for two models. The model with the larger, deeper void has a velocity shift deviating from that of the background model's at a level that is below the expected accuracy of the next generation ELT measurements. The velocity shift of the model with the smaller void deviates more.
\newline\newline
The effects of local overdensities have not been studied here but it is expected that the results will be qualitatively the same: Large, high overdensities are needed in order for the redshift drift to be affected at a level important for observations. See {\em e.g.} \cite{drift_LTB} for a study that includes effects of large local overdensities.
\newline\indent
Note lastly that the primary reason that voids need to be extremely large in order to affect observations is that ELT observations of the redshift drift are expected to be made only at large redshifts. Depending on the size of the uncertainty of low-redshift measurements, such measurements of the redshift drift may be affected by much smaller voids.

\subsection{Redshift drift in swiss-cheese models}
\begin{figure}[htb!]
\centering
\subfigure[]{
\includegraphics[scale = 0.8]{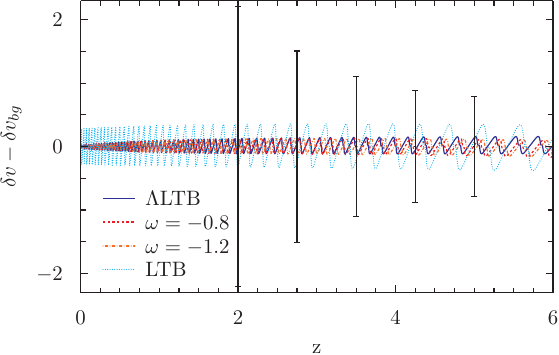}
}
\subfigure[]{
\includegraphics[scale = 0.8]{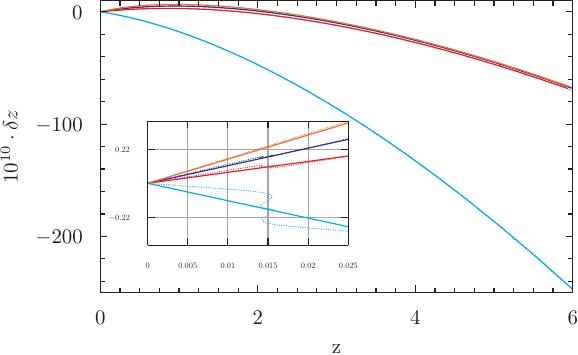}
}
\caption{Redshift drift and shift in the velocity shift in swiss-cheese models compared to their respective background models. The redshift drift of each of the four models is shown together with the redshift drift of the given background model. The fluctuations in the redshift drift due to the inhomogeneities are so small that they are only visible in a close-up. The figure to the right uses the same coloring scheme as that indicated in the figure to the left.}
\label{fig:cheese}
\end{figure}

\begin{figure}[htb!]
\centering
\subfigure[]{
\includegraphics[scale = 0.8]{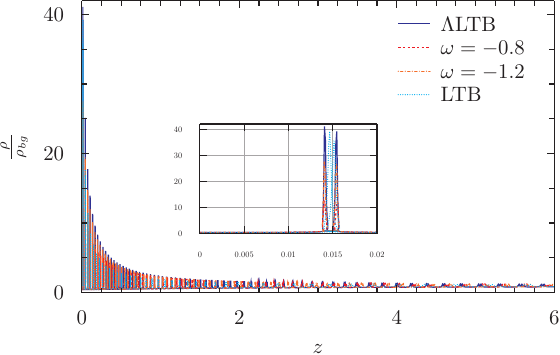}
}
\subfigure[]{
\includegraphics[scale = 0.8]{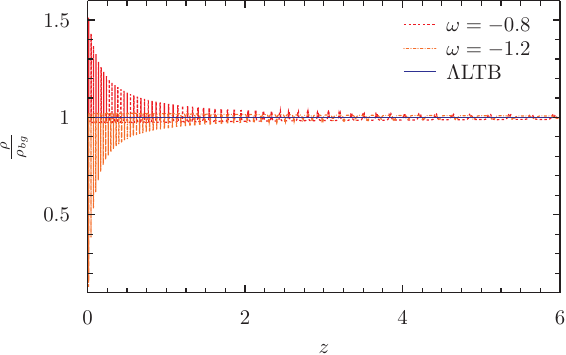}
}
\caption{Density along rays in swiss-cheese models. The legends indicate specifications of the background models (flat $\omega$CDM models with $\Omega_{m,0} = 0.3$ or $\Omega_{m,0} = 1$). The figure to the left shows fluctuations in the dark matter perturbations while the figure to the right shows fluctuations in the dark energy density. Fluctuations in the dark energy density are only present in the two models with $\omega = \omega_{de} = -0.8,-1.2$. The figure showing the dark matter density has a close-up of the low redshift region showing the first overdensity traveled through by the light rays.}
\label{fig:cheese_rho}
\end{figure}

Even disregarding large local structures, the redshift drift of a given object will not be exactly equal to the average redshift drift except in models that are exactly spatially homogeneous and isotropic. As discussed in section \ref{sec:caveats}, statistical fluctuations about the average value are expected to be small and negligible as long as observational results are averaged over a reasonable number of observed objects.
\newline\indent
As an illustration of the size that the fluctuations in the redshift drift can be expected to have, the redshift drift has been computed in four different swiss-cheese models. The models are constructed by combining single void models based on $k(r)$ given in equation (\ref{eq:k_comp}). The four models are based on four different backgrounds: an EdS model, a $\Lambda$CDM model and two $\omega$CDM models with $\omega = \omega_{de} = -0.8,-1.2$. The redshift drift has, in each model, been computed along a single ray that propagates radially through consecutive voids with present time radii of approximately 60Mpc. The results are shown in figure \ref{fig:cheese}. The fluctuations in the redshift drift compared to the background (and hence average) redshift drift are barely visible. The figure also shows the corresponding shift in the velocity shift compared to the background velocity shift. As seen, the fluctuations in the velocity shift are well below the accuracy expected from the next generation ELT observations. The results are based on $\delta t_0 = 30$ years.
\newline\indent
The density along the four rays is shown in figure \ref{fig:cheese_rho}. The dark matter densities along the four rays are not particularly distinguishable from each other and the point with the figure is merely to illustrate the order of the size of the density fluctuations. The dark energy fluctuations are clearly distinguishable from each other since $\omega<-1$ leads to fluctuations of the opposite sign of the dark matter fluctuations. Note that the density fluctuations of the dark energy components are much smaller than the perturbations in the dark matter density fields (see {\em e.g.} \cite{Mikko} for an elaboration of this point).
\newline\newline

The redshift drift and velocity shift fluctuations shown in figure \ref{fig:cheese} are due to inhomogeneities in the density and thus the geometry of spacetime, and not actually peculiar velocities. As shown in {\em e.g.} \cite{Sofie, Sofie2,Troels} (see also \cite{strong_to_weak, radial_cheese, Dallas_velocity}), inhomogeneities in spacetime can be interpreted in terms of peculiar velocities. The study conducted here can therefore equally be considered as illustrating effects of peculiar velocity fields as showing effects of an inhomogeneous geometry. However, studies based on {\em e.g.} perturbation theory indicate that peculiar acceleration is more important than effects of peculiar velocity. Since no actual peculiar velocity or acceleration fields have been computed here, the results presented above should be considered with caution. In addition, the results shown here are for specific swiss-cheese models. The fluctuations in the velocity shift may be significantly larger in other models. For example, the voids in the swiss-cheese models are only somewhat deep. By using equation (\ref{eq:k_uncomp}), other swiss-cheese models with deeper voids have been constructed. These models' fluctuations in $\delta v$ are also well below the expected ELT accuracy, in agreement with the results shown in figure \ref{fig:cheese}.


\end{document}